\documentclass{revtex4}
\usepackage{amsmath}

\begin{document}

\title{Effective 't Hooft-Polyakov monopoles from pure SU(3) gauge theory}
\author{Vladimir Dzhunushaliev}
\email{dzhun@hotmail.kg}
\affiliation{Dept. Phys. and Microel. Engineer., Kyrgyz-Russian
Slavic University, Bishkek, Kievskaya Str. 44, 720021, Kyrgyz
Republic}

\author{Douglas Singleton}
\email{dougs@csufresno.edu}
\affiliation{Physics Dept., CSU Fresno, 2345 East San Ramon Ave.
M/S 37 Fresno, CA 93740-8031, USA}

\date{\today}

\begin{abstract}
The well known topological monopoles originally investigated by 't Hooft
and Polyakov are known to arise in classical Yang-Mills-Higgs theory.
With a pure gauge theory it is known 
that the classical Yang-Mills field equation do not have such finite energy
configurations. Here we argue that such configurations may arise
in a semi-quantized Yang-Mills theory, where the original gauge group, $SU(3)$,
is reduced to a smaller gauge group, $SU(2)$, and with some combination
of  the coset fields of the $SU(3)$ to $SU(2)$ reduction acting as effective 
scalar fields. The procedure is called semi-quantized since some of the original
gauge fields are treated as quantum degrees of freedom, while others are postulated
to be effectively described as classical degrees of freedom. Some speculation
is offer on a possible connection between these monopole configurations and the 
confinement problem, and the nucleon spin puzzle.
\end{abstract}

\maketitle

\section{Introduction}

In two recent papers  \cite{vdsin1} \cite{vdsin2} it was pointed out that 
quantized SU(2) gauge theory 
in some approximation is equivalent to a U(1) gauge theory plus a scalar field, 
{\it i.e.} a pure SU(2) gauge theory reduces to a smaller 
U(1) abelian subgroup plus a symmetry breaking effective Higgs field. 
The postulate underlying this construction was that in some situations 
the SU(2) gauge fields can be decomposed into ordered and disordered 
phases. For the ordered phase the components of the SU(2) field have 
non-zero quantum average $\left\langle a^a_\mu \right\rangle \neq 0$, while the
disordered phase has a zero average $\left\langle A^m_\mu \right\rangle = 0$. 
Nevertheless, it is postulated that the condensate of  the disordered phase is non-zero
({\it i.e.} $\left\langle A^m_\mu A^n_\nu \right\rangle \neq 0$, ) so that it
therefore possesses a non-zero energy. Under these conditions 
the pure SU(2) gauge theory is equivalent to the Ginzburg - Landau theory 
interacting with the U(1) gauge field. 
\par 
The aim of this paper is to extend these results to the SU(3) gauge theory. 
There are interesting differences between the previous SU(2) $\rightarrow$ U(1) 
reduction and the present SU(3) $\rightarrow$ SU(2) reduction.
For the SU(2)$\rightarrow$U(1) case the gauge field, $a^a_\mu$, of the subgroup 
belongs to an Abelian subgroup U(1), and therefore does not have 
self-interaction terms like $(a^a_\mu a^{a\mu})^2$. In the 
SU(3)$\rightarrow$SU(2) case the gauge fields,  $a^a_\mu$ ($a=1,2,3$),  belong
to a non-Abelian subgroup SU(2), and these
gauge field do have self-interaction terms like $(a^a_\mu a^{a\mu})^2$. 
These terms are expected to change some results in comparison with the case 
investigated in Refs. \cite{vdsin2}. In particular in \cite{vdsin2} after the
reduction from SU(2) $\rightarrow$ U(1) we obtained an effective Ginzburg-Landau 
Lagrangian of the form which gives rise to Nielsen-Olesen flux tube solutions 
\cite{no}. In the present case we will find that the SU(3) $\rightarrow$ SU(2) 
reduction yields an effective Georgi-Glashow \cite{gg} Lagrangian which
has t' Hooft-Polyakov monopole solutions \cite{thp}. 
The mass of the monopole solutions is inversely proportional to the square of the
coupling constant and directly proportional to the mass scale or the mass of the 
relevant gauge bosons. The monopoles which occur in Grand Unified Theories like
SU(5) are therefore extremely massive since the gauge bosons have masses of order
$10 ^{15}$ GeV and the couplings constants are perturbatively small so that the 
monopoles would have masses greater than $10 ^{15}$ GeV. For the effective
Yang-Mills-Higgs Lagrangian derived here the coupling constant and mass scale 
would be that of SU(3) -- the coupling constant would be of
order $1$ and the mass scale would be of order $\Lambda \approx 200$ MeV. Thus
these monopole solutions would have a mass which would allow them to play
a physical role in low energy scale physics.
We offer some speculation that such solutions may play a role in the
nucleon spin puzzle \cite{emc}. The main point to be made is that while
pure Yang-Mills theories are proven not to have classical finite energy
solutions \cite{coleman} \cite{deser}, in the present case we perform a semi-quantization
of the system, and find that finite energy monopole solutions arise not from 
the classical theory, but from the quantized theory.

\section{SU(3)$\rightarrow$SU(2) + coset decomposition}

In this section the reduction of SU(3) to SU(2) is defined. 
We follow the conventions of Ref. \cite{kondo1}. 
Starting with the SU(3) gauge group with generators $T^B$, we define the SU(3)
gauge fields, $\mathcal{A}_\mu=\mathcal{A}^B_\mu T^B$. Let SU(2) be a subgroup
of SU(3) and SU(3)/SU(2) is a coset. Then the gauge field
$\mathcal{A}_\mu$ can be decomposed as
\begin{eqnarray}
  \mathcal{A}_\mu & = & \mathcal{A}^B_\mu T^B = a^a_\mu T^a + A^m_\mu T^m ,
\label{sec2-10a}\\
  a^a_\mu & \in & SU(2) \quad \text{and} \quad A^m_\mu \in SU(3)/SU(2) 
\label{sec2-10b}
\end{eqnarray}
where the indices $a,b,c \ldots $ belongs to the subgroup SU(2) and
$m,n, \ldots $to the coset SU(3)/SU(2); $B$ are SU(3) indices.
Based on this the field strength can be decomposed as
\begin{equation}
  \mathcal{F}^B_{\mu\nu} T^B = \mathcal{F}^a_{\mu\nu}T^a +
  \mathcal{F}^m_{\mu\nu}T^m
\label{sec2-20}
\end{equation}
where 
\begin{eqnarray}
  \mathcal{F}^a_{\mu\nu} & = & h^a_{\mu\nu} + \Phi^a_{\mu\nu}
  \; \; \; \in SU(2) ,
\label{sec2-30a}\\
  h^a_{\mu\nu} & = & \partial_\mu a^a_\nu - \partial_\nu a^a_\mu +
  g \epsilon^{abc}a^b_\mu a^c_\nu \; \; \; \in SU(2) ,
\label{sec2-30b}\\
  \Phi^a_{\mu\nu} & = & g f^{amn} A^m_\mu A^n_\nu \; \; \; \in SU(2) ,
\label{sec2-30c}\\
  \mathcal{F}^m_{\mu\nu} & = & F^m_{\mu\nu} + G^m_{\mu\nu} \; \; \;
  \in SU(3)/SU(2) ,
\label{sec2-30d}\\
  F^m_{\mu\nu} & = & \partial_\mu A^m_\nu - \partial_\nu A^m_\mu +
  g f^{mnp} A^n_\mu A^p_\nu \; \; \; \in SU(3)/SU(2) ,
\label{sec2-30e}\\
  G^m_{\mu\nu} & = & g f^{mnb}
  \left(
  A^n_\mu a^b_\nu - A^n_\nu a^b_\mu
  \right) \; \; \; \in SU(3)/SU(2)
\label{sec2-30f}  
\end{eqnarray}
where $f^{ABC}$ are the structure constants of SU(3), 
$\epsilon^{abc} = f^{abc}$ are the structure constants of SU(2). 
The SU(3) Yang-Mills field equations can be decomposed as
\begin{eqnarray}
  d_\nu \left( h^{a\mu\nu} +\Phi^{a\mu\nu} \right) & = &
  -g f^{amn} A^m_\nu
  \left(
  F^{n\mu\nu} + G^{n\mu\nu}
  \right) ,
\label{sec2-40a}\\
  D_\nu \left( F^{m\mu\nu} + G^{m\mu\nu} \right) & = &
  - g f^{mnb}
  \left[
  A^n_\nu \left( h ^{b\mu\nu} + \Phi^{b\mu\nu} \right) -
  a^b_\nu \left( F^{n\mu\nu} + G^{n\mu\nu} \right)
  \right]
\label{sec2-40b}
\end{eqnarray}
where $d_\nu [\cdots]^a = \partial_\nu [\cdots]^a +
f^{abc} a^b_\nu [\cdots]^c$ is the covariant derivative on the
subgroup SU(2) and
$D_\nu [\cdots]^m = \partial_\nu [\cdots]^m +
f^{mnp} A^n_\nu [\cdots]^p$

\section{Heisenberg quantization for QCD}

In the rest of the paper we will apply a modification of the Heisenberg
quantization technique to the system defined in the previous section.
In quantizing the classical system given in Eqs. (\ref{sec2-40a}) -
(\ref{sec2-40b}) via Heisenberg's method \cite{heis} one first replaces the
classical fields by field operators $a^a _{\mu} \rightarrow \hat{a}^a _\mu$
and $A^m_{\mu} \rightarrow \hat{A}^m_\mu$. This yields non-linear,
coupled, differential equations for the field operators. One then 
uses these equations to determine the 
expectation values for the field operators $\hat a ^a _\mu$ and
$\hat A^m_\mu$ ({\it e.g.} $\langle \hat a ^a _\mu \rangle$, where
$\langle \cdots \rangle = \langle Q | \cdots | Q \rangle$ and
$| Q \rangle$ is some quantum state). One can also use these
equations to determine the expectation values of operators
that are built up from the fundamental operators $\hat a^a_\mu$
and $\hat A^m_\mu$. For example, the ``electric'' field
operator, $\hat E^a_z = \partial _0 \hat a^a_z - \partial _z \hat a^a_0$
giving the expectation $\langle \hat E^a_z \rangle$.
The simple gauge field expectation values,
$\langle \mathcal{A}_\mu (x) \rangle$, are obtained by
taking the expectation of the operator version of Eqs. \eqref{sec2-40a} \eqref{sec2-40b} 
with respect to some quantum state $| Q \rangle$.
One problem in using these equations to obtain expectation
values like $\langle A^m_\mu \rangle$, is that these equations
involve not only powers or derivatives of $\langle A^m_\mu \rangle$
({\it i.e.} terms like $\partial_\alpha \langle A^m_\mu \rangle$ or
$\partial_\alpha \partial_\beta \langle A^m_\mu \rangle$)
but also contain terms like $\mathcal{G}^{mn}_{\mu\nu} =
\langle A^m_\mu A^n_\nu \rangle$. Starting with the
operator version of Eqs. (\ref{sec2-40a})--
(\ref{sec2-40b}) one can generate an operator differential
equation for the product $\hat A^m_\mu \hat A^n_\nu$ thus allowing
the determination of the Green's function $\mathcal{G}^{mn}_{\mu\nu}$.
However this equation will in turn contain other, higher order
Green's functions. Repeating these steps leads to an infinite set
of equations connecting Green's functions of ever increasing
order. This procedure is very similar to the field correlators approach
in QCD (for a review, see \cite{giacomo}). In Ref. \cite{simonov} 
a set of self coupled equations for such field correlators is given. 
This construction, leading to an infinite set of coupled,
differential equations, does not have an exact, analytical solution
and so must be handled using some approximation.
\par
Operators are only well determined if there is a Hilbert space of quantum
states. Thus we need to ask about the definition of the quantum states
$| Q \rangle$ in the above construction. The resolution to this
problem is as follows: There is an one-to-one correspondence
between a given quantum state $| Q \rangle$ and the infinite set
of quantum expectation values over any product of field operators,
$\mathcal{G}^{mn \cdots}_{\mu\nu \cdots}(x_1, x_2 \ldots) =
\langle Q | A^m_\mu (x_1) A^n_\nu (x_2) \ldots
| Q \rangle$. So if all the Green's functions
-- $\mathcal{G}^{mn \cdots}_{\mu\nu \cdots}(x_1, x_2 \ldots)$ --
are known then the quantum states, $| Q \rangle$ are known,
\textit{i.e.} the action of $| Q \rangle$ on any product
of field operators $\hat A^m_\mu (x_1) \hat A^n_\nu (x_2) \ldots$
is known. The Green's functions are determined from the above,
infinite set of equations (following Heisenberg's idea).
\par
Another problem associated with products of field operators
like $\hat A^m_\mu (x) \hat A^n_\nu (x)$ is that the two operators occur at the
same point. For \textit{non-interacting} field it is well
known that such products have a singularity. In this paper
we are considering \textit{interacting} fields so it is
not known if a singularity would arise for such products
of operators evaluated at the same point. Physically
it is hypothesized that there are situations in interacting
field theories where these singularities do not occur
({\it e.g.} for flux tubes in Abelian or non-Abelian theory
quantities like the ``electric'' field inside the tube,
$\langle E^a_z \rangle < \infty$, and energy density
$\varepsilon (x) = \langle (E^a_z)^2 \rangle < \infty$ are
nonsingular). Here we take as an assumption that such singularities
do not occur.

\section{Basic assumptions}

It is evident that the full and exact quantization is impossible in this 
case. Thus we have to look for some simplification in order to obtain
equations which can be analyzed. Our basic aim is cut off the infinite 
equations set using some simplifying assumptions. For this purpose we have 
to have ans\"atzen for the following 2 and 4-points Green's functions : 
$\left\langle A^m_\mu(y) A^n_\nu(x) \right\rangle$, 
$\left\langle a^a_\alpha (x) a^b_\beta (y) A^m_\mu(z) A^n_\nu(u) \right\rangle$ and 
$\left\langle A^m_\alpha (x) A^n_\beta (y) A^p_\mu(z) A^q_\nu(u) \right\rangle$. 
At first we assume that there are two phases 
\begin{enumerate}
\item The gauge field components $a^a_\mu$ (a=1,2,3 $a^a_\mu \in SU(2)$) 
      belonging to the small subgroup SU(2) are in an ordered phase. Mathematically
      this means that 
\begin{equation}
  \left\langle a^a_\mu (x) \right\rangle  =(a^a _{\mu} (x))_{cl}.
\label{sec4-10}
\end{equation}
      The subscript means that this is the classical field. Thus we are 
      treating these components as effectively classical gauge fields 
      in the first approximation. 
\item The gauge field components $A^m_\mu$ (m=4,5, ... , 8 and 
      $A^m_\mu \in SU(3)/SU(2)$) belonging to the coset SU(3)/SU(2) are in 
      a disordered phase (or in other words - a condensate), but have 
      non-zero energy. In mathematical terms this means that 
\begin{equation}
  \left\langle A^m_\mu (x) \right\rangle = 0, 
  \quad \text{but} \quad 
  \left\langle A^m_\mu (x) A^n_\nu (x) \right\rangle \neq 0 .
\label{sec4-20}
\end{equation}
      Later we will postulate a specific, and physically reasonable form for the non-zero term. 
\item There is not correlation between ordered (classical) and disordered (quantum) phases 
\begin{equation}
  \left\langle f(a^a_\mu) g(A^m_\nu) \right\rangle =
  f(a^a_\mu)  \left\langle g(A^m_\mu) \right\rangle
\label{sec4-25}
\end{equation}
Later we will give a specific form for this correlation in a 4-point Green's function.
\end{enumerate}

\section{Derivation of an effective Lagrangian}

Our quantization procedure will derivate from the Heisenberg method in that we will take the expectation
of the Lagrangian rather than for the equations of motions. Thus we will obtain an effective Lagrangian rather
than approximate equations of motion. The Lagrangian we obtain from the original SU(3) pure gauge
theory is an effective SU(2) Yang-Mills-Higgs system which has monopole solutions. The averaged Lagrangian is 
\begin{equation}
  \mathcal{L} = - \frac{1}{4}\left\langle \mathcal{F}^A_{\mu\nu} \mathcal{F}^{A\mu\nu}
  \right\rangle = 
- \frac{1}{4} \left( \left\langle \mathcal{F}^a_{\mu\nu} \mathcal{F}^{a\mu\nu} \right\rangle + 
  \left\langle \mathcal{F}^m_{\mu\nu} \mathcal{F}^{m\mu\nu} \right\rangle \right) 
\label{sec5-10}
\end{equation}
here $\mathcal{F}^{a\mu\nu}$ and $\mathcal{F}^{m\mu\nu}$ are defined by 
equations \eqref{sec2-30a}-\eqref{sec2-30f}. 

\subsection{Calculation of 
$\left\langle \mathcal{F}^a_{\mu\nu} \mathcal{F}^{a\mu\nu} \right\rangle$}

We begin by calculating the first term on the r.h.s. of equation 
\eqref{sec5-10}
\begin{equation}
  \left\langle \mathcal{F}^a_{\mu\nu} \mathcal{F}^{a\mu\nu} \right\rangle = 
  \left\langle h^a_{\mu\nu} h^{a\mu\nu} \right\rangle + 
  2\left\langle h^a_{\mu\nu} \Phi^{a\mu\nu} \right\rangle + 
  \left\langle \Phi^a_{\mu\nu} \Phi^{a\mu\nu} \right\rangle . 
\label{sec5-20}
\end{equation}
Immediately we see that the first term on the r.h.s. of this equation is 
SU(2) Lagrangian as we assume that $a^a_\mu$ and $h^a_{\mu\nu}$ are the 
classical quantities and consequently 
\begin{equation}
  \left\langle h^a_{\mu\nu} h^{a\mu\nu} \right\rangle \approx 
  h^a_{\mu\nu} h^{a\mu\nu} . 
\label{sec5-30}
\end{equation}
The second term in equation \eqref{sec5-10} is
\begin{equation}
  \left\langle h^a_{\mu\nu} \Phi^{a\mu\nu} \right\rangle = 
  g f^{amn} 
  \left\langle  
  \left(
  \partial_\mu a^a_\nu - \partial_\nu a^a_\mu
  \right)
  A^{m\mu} A^{n\nu}
  \right\rangle + 
  g f^{abc} f^{amn} 
  \left\langle
  a^b_\mu a^c_\nu A^{m\mu} A^{n\nu}
  \right\rangle . 
\label{sec5-40}
\end{equation}
Using assumptions 1 and 3 from the previous section these terms become
\begin{equation}
  \left\langle a^a_\alpha (x) A^m_\mu (y) A^n_\nu (z) \right\rangle = 
  a^a_\alpha (x) \left\langle A^m_\mu (y) A^n_\nu (z) \right\rangle =  
  \eta_{\mu\nu} a^a_\alpha (x) \mathcal{G}^{mn} (y,z)
\label{sec5-40a}
\end{equation}
and
\begin{equation}
  \left\langle a^a_\alpha (x) a^b_\beta (y) A^m_\mu (z) A^n_\nu (u) \right\rangle 
  = a^a_\alpha (x) a^b_\beta (y) \eta_{\mu\nu} \mathcal{G}^{mn} (z,u)
\label{sec5-40b}
\end{equation} 
The function $\mathcal{G}^{mn} (x,y)$ is the 2-point correlator (Green's
function) for the disordered phase. Because of the bosonic character of the 
coset gauge fields $\mathcal{G}^{mn} (x,y)$ must be symmetric under exchange
of these fields. Also by assumption 2 of the last section this expectation should
be non-zero. We take the form for this 2-point correlator to be
\begin{equation}
  \left\langle A^m_\mu (y) A^n_\nu (x) \right\rangle = 
  - \frac{1}{3}\eta_{\mu\nu} f^{mpb} f^{npc} 
  \phi^b (y) \phi^c (x) = 
  - \eta_{\mu\nu} \mathcal{G}^{mn} (y,x)
\label{sec5-45}
\end{equation}
with
\begin{equation}
  \mathcal{G}^{mn} (y,x) = \frac{1}{3} f^{mpb} f^{npc} \phi^b (y) \phi^c (x)
\label{sec5-45a}
\end{equation}
here $\phi^a$ is a real SU(2) triplet scalar fields. Thus we have replaced the
coset gauge fields by an effective scalar field, which will be the scalar field in our
effective SU(2)-scalar system. The factor of $- \frac{1}{3}$ is introduced so that the
effective scalar field, $\phi$, will have the correct coefficent 
for the kinetic energy term. If we were to take the scalar field to be constant
($\phi ^a (x) \approx const.$) then \eqref{sec5-45} and \eqref{sec5-45a} would 
represent an effective mass-like, condensation term  of the coset gauge fields.
With this we find that the middle term vanishes
\begin{equation}
  \left\langle h^a_{\mu\nu} \Phi^{a\mu\nu} \right\rangle = g \eta^{\mu\nu} 
  \left( f^{amn} \left(\partial_\mu a^a_\nu (x) - \partial_\nu a^a_\mu (x) \right) 
  \mathcal{G}^{mn} (x,x) + 
  g f^{amn} f^{abc} a^b_\mu (x) a^c_\nu (x) \mathcal{G}^{mn} (x,x) 
  \right) 
  = 0 
\label{sec5-50}
\end{equation}
The last term which is quartic in the coset gauge fields will be considered at
the end. Up to this point the SU(2) part of the Lagrangian is 
\begin{equation}
  \left\langle \mathcal{F}^a_{\mu\nu} \mathcal{F}^{a\mu\nu} \right\rangle = 
  \left\langle h^a_{\mu\nu} h^{a\mu\nu} \right\rangle + 
  g^2 f^{anp} f^{an'p'} 
  \left\langle
  A^n_\mu A^p_\nu A^{n'\mu} A^{p'\nu} 
  \right\rangle . 
\label{sec5-60}
\end{equation}

\subsection{Calculation of 
$\left\langle \mathcal{F}^m_{\mu\nu} \mathcal{F}^{m\mu\nu} \right\rangle$}

Next we work on the coset part of the Lagrangian
\begin{equation}
\begin{split}
  \left\langle \mathcal{F}^m_{\mu\nu} \mathcal{F}^{m\mu\nu} \right\rangle = 
  \left\langle
   \left[
    \left(
    \partial_\mu A^m_\nu - g f^{mnb} A^n_\nu a^b_\mu
    \right) - 
    \left(
    \partial_\nu A^m_\mu - g f^{mnb} A^n_\mu a^b_\nu
    \right) + 
    g f^{mnp} A^n_\mu A^p_\nu
   \right]^2 
  \right\rangle &= \\
  2\left\langle
    \left(
    \partial_\mu A^m_\nu - g f^{mnb} A^n_\nu a^b_\mu
    \right) ^2 
    \right\rangle - 
  2\left\langle
    \left(
    \partial_\mu A^m_\nu - g f^{mnb} A^n_\nu a^b_\mu
    \right)
    \left(
    \partial^\nu A^{m\mu} - g f^{mn'b'} A^{n'\mu} a^{b'\nu}
    \right)
   \right\rangle + \\
   4g \left\langle
    \left(
    \partial_\mu A^m_\nu - g f^{bmn'} A^{n'}_\nu a^b_\mu
    \right)
    f^{mnp} A^{n\mu} A^{p\nu}
   \right\rangle + 
   g^2 f^{mnp} f^{mn'p'} 
   \left\langle
   A^n_\mu A^p_\nu A^{n'\mu} A^{p'\nu}
   \right\rangle .
\end{split}
\label{sec5-100}
\end{equation}
First we calculate 
\begin{equation}
\begin{split}
  2 \left\langle \left( \partial_\mu A^m_\nu \right)^2 \right\rangle = 
  2 \left\langle
  \partial_{y\mu} A^m_\nu (y) \partial^\mu_x A^{m\nu} (x)
  \right\rangle \Bigr |_{y=x} = 
  2 \partial_{y\mu} \partial^\mu_x 
  \left\langle A^m_\nu (y) A^{m\nu} (x) \right\rangle \Bigr |_{y=x} &= \\
  - \frac{2}{3} \eta^\nu_\nu f^{mpb} f^{mpc} \partial_\mu \phi^b \partial^\mu \phi^c = 
  - \frac{8}{3} \partial_\mu \phi^a \partial^\mu \phi^a & .
\label{sec5-110}
\end{split}
\end{equation}
Analogously 
\begin{equation}
  -2 \left\langle
  \partial_\mu A^m_\nu (y) \partial^\nu A^{m\mu} (x) 
  \right\rangle \Bigr |_{y=x} = 
  \frac{2}{3} \eta^\mu_\nu f^{mpb} f^{mpc} \partial_\mu \phi^b \partial^\nu \phi^c = 
  \frac{2}{3} \partial_\mu \phi^a \partial^\mu \phi^a . 
\label{sec5-120}
\end{equation}
The next term is 
\begin{equation}
\begin{split}
  - 4 g  \left\langle
   \left(
   \partial_\mu A^m_\nu 
   \right) f^{amn} A^{n\nu} a^{a\mu} 
  \right\rangle = 
  - 4 g f^{amn} a^{a\mu} 
  \left\langle
   \left(
   \partial_\mu A^m_\nu  
   \right) A^{n\nu} 
  \right\rangle = 
  - 4 g f^{amn} a^{a\mu} (x) 
  \left\langle
   \left(
   \partial_{y\mu} A^m_\nu (y)  
   \right) A^{n\nu} (x) 
  \right\rangle \Bigr |_{y=x} & = \\ 
  \frac{4}{3} g \eta^\nu_\nu f^{amn} a^{a\mu} f^{mpb} f^{npc} 
  \partial_\mu \phi^b \phi^c = 
  - \frac{16}{3} g a^{a\mu} \left( f^{amn} f^{cnp} f^{bpm} \right) 
  \partial_\mu \phi^b \phi^c = 
   \frac{8}{3} g \epsilon^{abc} a^{a\mu} \partial_\mu \phi^b \phi^c &
\end{split}
\label{sec5-130}
\end{equation}
using $f^{amn}f^{cnp} f^{bpm} = \frac{1}{2} \epsilon^{acb}$. Analogously 
\begin{equation}
4 g  \left\langle
   f^{amn} A^n_\nu (y) a^a_\mu (y) 
   \left(
   \partial^\nu_x A^{m\mu} (x)
   \right) 
  \right\rangle \Bigr |_{y=x} = 
  \frac{4}{3} g \left( f^{amn} f^{bnp} f^{cpm} \right) a^{a\mu} 
  \phi^b \partial^\mu \phi^c = 
  - \frac{2}{3} g \epsilon^{abc} a^{a\mu} \partial_\mu \phi^b \phi^c . 
\label{sec5-140}
\end{equation}
Using \eqref{sec5-45} the next term is 
\begin{equation}
\begin{split}
  2 g^2 \left\langle
  f^{dmn} A^n_\nu a^d_\mu f^{d'mn'} A^{n'\nu} a^{d'\mu} 
  \right\rangle = 
 2 g^2 f^{dmn} a^d_\mu f^{d'mn'}  a^{d'\mu} 
\left\langle  A^n_\nu  A^{n'\nu} \right\rangle &= \\
  - \frac{2}{3} \eta^\nu_\nu 
  \left( f^{dmn} f^{d'mn'} f^{npb} f^{n'pc} \right)
  \left( a^d_\mu a^{d'\mu} \phi^b \phi^c \right) = 
  - \frac{8}{3} g^2 E^{d'dbc}  a^d_\mu a^{d'\mu} \phi^b \phi^c 
   &
\end{split}
\label{sec3-160}
\end{equation}
here $E^{d'dbc} = f^{d'n'm} f^{dmn} f^{bnp} f^{cpn'}$ and its components are 

\begin{equation}
\begin{split}
  E^{aaaa} = E^{1111} = E^{2222} = E^{3333} = \frac{1}{4} & \\
  E^{aabb} = E^{1122} = E^{1133} = E^{2211} = 
  E^{2233} = E^{3311} = E^{3322} = \frac{1}{4} & \\
  E^{abab} = -E^{1212} = E^{1221} = -E^{1313} = 
  E^{1331} = -E^{2121} = E^{2112} = 
  -E^{1313} = E^{3113} = -E^{3232} = E^{3223} = \frac{1}{4} .&
\end{split}
\label{sec3-170}
\end{equation}
We now note that $E^{d'dbc} a^d_\mu a^{d'\mu} \phi^b \phi^c 
 = \frac{1}{4}( \epsilon ^{abc} \epsilon^{ab'c'} a_{\mu} ^b \phi ^c a^{b' \mu} 
\phi ^{c'} + a_{\mu} ^b \phi ^b a^{c \mu} \phi ^c)$. Thus \eqref{sec3-160} becomes
\begin{equation}
  2 g^2 \left\langle
  f^{dmn} A^n_\nu a^d_\mu f^{d'mn'} A^{n'\nu} a^{d'\mu} 
  \right\rangle = 
- \frac{2 g^2}{3} (\epsilon ^{abc} \epsilon^{ab'c'} a_{\mu} ^b \phi ^c a^{b' \mu} 
\phi ^{c'} + a_{\mu} ^b \phi ^b a^{c \mu} \phi ^c ) 
\label{sec3-180}
\end{equation}
Analogously
\begin{equation}
-2   g^2 \left\langle
  f^{dmn} A^n_\nu a^d_\mu f^{d'mn'} A^{n'\mu} a^{d'\nu} 
  \right\rangle = 
  \frac{g^2}{6} (\epsilon^{abc} \epsilon^{ab'c'} a^b_\mu \phi^c a^{b'\mu} \phi^{c'}
 + a_{\mu} ^b \phi ^b a^{c \mu} \phi ^c )
\label{sec3-190}
\end{equation}
Finally,  there are the terms that involve three coset fields ({\it e.g.} 
$ \langle f^{mnp} ( \partial_\mu A^m_\nu A^{n\mu} A^{p\nu} ) \rangle$
and $ f^{bmn'} f^{mnp} a^b_\mu \langle A^{n'}_\nu  A^{n\mu} A^{p\nu}  \rangle$) 
The term involving the derivative is
\begin{equation}
f^{mnp} \langle  (\partial _{y \mu} A^m _\nu (y)) A^{n\mu} (x) A^{p\nu} (z) \rangle 
\label{sec3-185a}
\end{equation}
Since the gauge fields must be symmetric under exchange, and because of the 
antisymmetry of the of $f^{mnp}$ this term vanishes. Next the terms involving 
three coset fields and one $SU(2)$ field we will approximate as
\begin{equation}
f^{bmn'} f^{mnp} a^b_\mu \langle A^{n'}_\nu  A^{n\mu} A^{p\nu}  \rangle \approx
\frac{1}{3} f^{bmn'} f^{mnp} a^b_\mu \left( \langle A^{n'}_\nu 
\rangle \langle A^{n\mu} A^{p\nu}  \rangle 
+\langle A^{n'}_\nu  A^{n\mu} \rangle \langle A^{p\nu}  \rangle 
+\langle A^{n'}_\nu  A^{p\nu} \rangle \langle A^{n\mu} \rangle \right) 
\label{sec3-185b}
\end{equation}
By the second assumption in the previous section, $\langle A ^m _\mu (x) \rangle =0$, this
term also vanishes. Thus 
\begin{equation}
\begin{split}
  \left\langle \mathcal{F}^m_{\mu\nu} \mathcal{F}^{m\mu\nu} \right\rangle
  = - 2 \partial_\mu \phi^{a} \partial^\mu \phi^a + 
  2 g \epsilon^{abc} a^{a\mu} \partial_\mu \phi^{b} \phi^c  -
  \frac{g^2}{2} \epsilon^{abc} \epsilon^{ab'c'} a^b_\mu \phi^c a^{b'\mu} \phi^{c'} 
& \; \\
 - \frac{g^2}{2} a_{\mu} ^b \phi ^b a^{c \mu} \phi ^c +  
  g^2 f^{mnp} f^{mn'p'} 
  \left\langle A^n_\mu A^p_\nu A^{n'\mu} A^{p'\nu} \right\rangle &= \\
 - 2 \left(
  \partial_\mu \phi^a + \frac{g}{2} \epsilon^{abc} a^b_\mu \phi^c
  \right)^2 - 
  \frac{g^2}{2} a_{\mu} ^b \phi ^b a^{c \mu} \phi ^c + 
  g^2 f^{mnp} f^{mn'p'} 
  \left\langle A^n_\mu A^p_\nu A^{n'\mu} A^{p'\nu} \right\rangle .
\end{split}
\label{sec200}
\end{equation}
The full averaged Lagrangian is 
\begin{equation}
- \frac{1}{4}  \left\langle \mathcal{F}^A_{\mu\nu} \mathcal{F}^{A\mu\nu} \right\rangle = 
  - \frac{1}{4}  h^a_{\mu\nu} h^{a\mu\nu} + 
  \frac{1}{2}   \left(
  \partial_\mu \phi^a + \frac{g}{2} \epsilon^{abc} a^b_\mu \phi^c
  \right)^2 + 
  \frac{g^2}{2} a_{\mu} ^b \phi ^b a^{c \mu} \phi ^c - 
  \frac{1}{4}  g^2 f^{Anp} f^{An'p'} 
  \left\langle A^n_\mu A^p_\nu A^{n'\mu} A^{p'\nu} \right\rangle .
\label{sec3-210}
\end{equation}
where we have collected the quartic terms from eq. \eqref{sec5-60} and \eqref{sec200}
together into 
$f^{Anp} f^{An'p'} \left\langle A^n_\mu A^p_\nu A^{n'\mu} A^{p'\nu} \right\rangle$. 

\subsection{The quartic term}

In this section we show that the quartic term -- $  f^{Anp} f^{An'p'} 
  \left\langle A^n_\mu A^p_\nu A^{n'\mu} A^{p'\nu} \right\rangle $ --
from eq. \eqref{sec3-210} becomes an effective $\lambda \phi ^4$
interaction term for the effective scalar field introduced in  eq. \eqref{sec5-45}.
Just as in eq. \eqref{sec5-45} where a quadratic gauge field expression was
replaced by a quadratic effective scalar field expression, here we replace
the quartic gauge field term by a quartic scalar field term 
\begin{equation}
  \left\langle
  A^m_\alpha (x) A^n_\beta (y) A^p_\mu (z) A^q_\nu (u) 
  \right\rangle = 
  \left(
  E^{mnpq}_{1,abcd} \eta_{\alpha\beta} \eta_{\mu\nu} + 
  E^{mpnq}_{2,abcd} \eta_{\alpha\mu} \eta_{\beta\nu} + 
  E^{mqnp}_{3,abcd} \eta_{\alpha\nu} \eta_{\beta\mu}
  \right) 
  \phi^a (x) \phi^b(y) \phi^c (z) \phi^d(u) 
\label{sec3-240}
\end{equation}
here $E^{mnpq}_{1,abcd}, E^{mpnq}_{2,abcd}, E^{mqnp}_{3,abcd}$ are constants. 
Because of the bosonic character of the gauge fields in \eqref{sec3-240} the 
indices of these constants in conjunction with the indices of the 
$\eta _{\alpha \beta}$'s must reflect symmetry under exchange of
the fields.  The simplest choice that satisfies this requirement is 
\begin{equation}
  \left\langle
  A^m_\alpha (x) A^n_\beta (y) A^p_\mu (z) A^q_\nu (u) 
  \right\rangle = 
  \left(
  \delta^{mn}\delta^{pq} \eta_{\alpha\beta} \eta_{\mu\nu} + 
  \delta^{mp}\delta^{nq} \eta_{\alpha\mu} \eta_{\beta\nu} + 
  \delta^{mq}\delta^{np} \eta_{\alpha\nu} \eta_{\beta\mu}
  \right) e_{abcd}
  \phi^a (x) \phi^b(y) \phi^c (z) \phi^d(u) 
\label{sec3-245}
\end{equation}
This choice of taking the constants from eq. \eqref{sec3-240} to be products 
of Kronecker deltas and fixing $a=b=c=d$  for the lower indices, satisfies 
the bosonic character requirement for the gauge fields, and is equivalent 
to the reduction used for the quartic term in \cite{vdsin2}. 
Evaluating eq. \eqref{sec3-245} at one spacetime point ({\it i.e.} $x=y=z=u$) 
and contracting the indices to conform to quartic term in eq. \eqref{sec3-210} 
gives
\begin{equation}
  \left\langle
  A^m_\alpha (x) A^n_\beta (x) A^p_\mu (x) A^q_\nu (x) 
  \right\rangle = 
  \left(
  \delta^{mn}\delta^{pq}  \eta_{\mu\nu} \eta^{\mu\nu} + 
  \delta^{mp}\delta^{nq} \eta_{\mu} ^{\mu} \eta_{\nu} ^{\nu} + 
  \delta^{mq}\delta^{np} \eta_{\mu} ^{\nu} \eta_{\nu} ^{\mu}
  \right) 
  \left( \phi^a (x) \phi^a(x) \right)^2 
\label{sec3-250}
\end{equation}
where the constants $e_{abcd}$ are chosen that at the point  $x=y=z=u$
\begin{equation}\label{sec3-255}
    e_{abcd} = \frac{1}{3}
    \left(
    \delta_{ab} \delta_{cd} + \delta_{ac} \delta_{bd} + 
    \delta_{ad} \delta_{bc}
    \right) .
\end{equation} 
This expression can be further simplified to
\begin{equation}
  \left\langle
  A^m_\mu A^n_\nu A^{p\mu} A^{q\nu} 
  \right\rangle = 
  \left(
  4 \delta^{mn}\delta^{pq}   + 
  16 \delta^{mp}\delta^{nq}  + 
  4  \delta^{mq}\delta^{np}
  \right) 
  \left( \phi^a (x) \phi^a(x) \right)^2 .
\label{sec3-260}
\end{equation}
Substituting this into the original quartic term of eq. \eqref{sec3-210} yields
\begin{equation}
\begin{split}
  f^{Anp} f^{An'p'} 
  \left\langle A^n_\mu A^p_\nu A^{n'\mu} A^{p'\nu} \right\rangle = 
  (4 f^{Ann} f^{An'n'} + 16 f^{Anp} f^{Anp} + 4 f^{Anp} f^{Apn}) 
  \left( \phi^a (x) \phi^a(x) \right)^2  & \; \\
= 12  f^{Anp} f^{Anp}  \left( \phi^a (x) \phi^a(x) \right)^2 
\label{sec3-270}
\end{split}
\end{equation}
where the antisymmetry property of the structure constants has been used. 
Using the explicit expression for the structure constants 
($f^{123} =1  ; \; \; f^{147}=f^{246}=f^{257} =
f^{345}=f^{516}=f^{637} = \frac{1}{2} ; \; \; f^{458} =f^{678} =\frac{\sqrt{3}}{2}$ 
plus those related to these by permutations), and recalling that the index 
$A$ runs from $1-8$ while the indices $n, p$ run from $4-8$ one can show that 
$f^{Anp} f^{Anp}= 12$ ($f^{123}$ 
and related constants do not contribute to this expression). Combining 
these results transforms the quartic term in eq. \eqref{sec3-210} as
\begin{equation}
 \frac{1}{4}  g^2 f^{Anp} f^{An'p'} 
  \left\langle A^n_\mu A^p_\nu A^{n'\mu} A^{p'\nu} \right\rangle =
36 g^2 \left( \phi^a (x) \phi^a(x) \right)^2 
\equiv \lambda \left( \phi^a (x) \phi^a(x) \right)^2 
\label{sec3-280}
\end{equation}
This has transformed the quartic gauge field term of the coset fields into a quartic
interaction term for the effective scalar field. Substituting this result back into the
averaged Lagrangian of eq. \eqref{sec3-210} we find
\begin{equation}
- \frac{1}{4}  \left\langle \mathcal{F}^A_{\mu\nu} \mathcal{F}^{A\mu\nu} \right\rangle = 
  - \frac{1}{4}  h^a_{\mu\nu} h^{a\mu\nu} + 
  \frac{1}{2}   \left(
  \partial_\mu \phi^a + \frac{g}{2} \epsilon^{abc} a^b_\mu \phi^c
  \right)^2 + 
  \frac{g^2}{2} a_{\mu} ^b \phi ^b a^{c \mu} \phi ^c - 
\lambda \left( \phi^a (x) \phi^a(x) \right)^2 
\label{sec3-290}
\end{equation}
The original pure SU(3) gauge theory has been transformed into an SU(2) gauge 
theory coupled to an effective triplet scalar field. This is similar to the 
Georgi-Glashow \cite{gg} Lagrangian except for the presence of the term  
$\frac{g^2}{2} a_{\mu} ^b \phi ^b a^{c \mu} \phi ^c$ and 
the absence of a negative mass term for $\phi ^a$ of the form 
$m^2 \phi^a (x) \phi^a(x)$. 

The Georgi-Glashow Lagrangian is known to have topological monopole solutions 
\cite{thp} which have the form
\begin{equation}
\phi ^a =  \frac{x^a f(r)}{g r^2} \; \;  ; \; \; a^a_0 = 0 \; \;  ; \; \; 
a^a _i = \frac{\epsilon _{aib} x^b [1- h(r)]}{g r^2}
\label{sec3-300}
\end{equation}
where $f(r)$ and $h(r)$ are functions determined by the field equations. 
For this form of the scalar and SU(2) gauge fields the term, 
$\frac{g^2}{2} a_{\mu} ^b \phi ^b a^{c \mu} \phi ^c$,
vanishes from the Lagrangian in eq. \eqref{sec3-290} by the antisymmetry 
of $\epsilon _{aib}$ and the symmetry of $x^a x^b$. Thus for the monopole 
ansatz of eq. \eqref{sec3-300} the Lagrangian in \eqref{sec3-290} becomes 
equivalent to the Georgi-Glashow Lagrangian minus only the mass term for 
the scalar field. 

In the present work we simply postulate that the effective scalar field develops 
a negative mass term of the form $-\frac{m^2}{2} (\phi ^a \phi ^a )$ which is 
added by hand to the Lagrangian of
\eqref{sec3-290} to yield
\begin{equation}
- \frac{1}{4}  h^a_{\mu\nu} h^{a\mu\nu} + 
  \frac{1}{2}   \left(
  \partial_\mu \phi^a + \frac{g}{2} \epsilon^{abc} a^b_\mu \phi^c
  \right)^2 + \frac{m^2}{2} (\phi ^a \phi ^a ) 
  - \lambda \left( \phi^a (x) \phi^a(x) \right)^2 +
\frac{g^2}{2} a_{\mu} ^b \phi ^b a^{c \mu} \phi ^c 
\label{sec3-310}
\end{equation}
The scalar field now has the standard symmetry breaking form and this effective 
Lagrangian has finite energy 't Hooft-Polyakov solutions (the last term should 
not alter the monopole construction since it vanishes under the ansatz in 
\eqref{sec3-300}). 

Our final result given in \eqref{sec3-310} depends on several crucial assumptions
({\it e.g.} the existence of the negative mass term $-\frac{m^2}{2} (\phi ^a \phi ^a )$).
In the next section we make some remarks and discuss possible motivations for some 
of the major assumptions.  

\section{Remarks/speculations about mass $m^2(\phi^a)^2$ and 
$a^b_\mu \phi^b a^{c\mu} \phi^c$ terms}

First we would like to emphasize that the calculations presented here 
are nonperturbative in the following sense. It is well known that perturbative 
techniques do not work in QCD. The degrees of freedom dealt with in this
paper have been split into two phases. The first phase is an order phase 
which is treated as an effectively classical degree of freedom. Perturbations around 
corresponding solutions probably can be calculated perturbatively   
using something like Feynman diagram technique. The second phase 
is a purely quantum, nonperturbative degree of freedom. In the first 
approximation these degrees of freedom can be calculated with our simplifications presented above. 
We can assume that as above in the first case the perturbations around the
correlators (Green's function \eqref{sec5-45a}) can be calculated using 
something like Feynman diagram techniques. 
\par 
In counting the degrees of freedom there is an apparent mismatch between the
final and initial degrees of freedom. The 
averaged Lagrangian \eqref{sec3-310} has $3 \times 2 = 6$ 
(for $a^b_\mu$) + 3 (for $\phi ^a$) = 9 degrees of freedom, while the initial SU(3), 
QCD Lagrangian has $8 \times 2 = 16$ degrees of freedom. Thus there is an
apparent shortfall of 16 - 9 = 7 degrees of freedom. This shortfall occurs in
eqs. \eqref{sec5-45} \eqref{sec5-45a} where $5 \times 2 =10$ degrees of
freedom from the coset gauge fields, $A^m _{\mu}$, are put into 
$3$ degrees of freedom, $\phi ^a (x)$. Our postulate is that QCD
has nonperturbative and perturbative degrees of freedom. The initial SU(3) 
Lagrangian contains both kinds of degrees of freedom. 
The final effective Lagrangian in eq. \eqref{sec3-310}
contains the nonperturbative degrees of freedom -- the SU(2) gauge fields
$a_{\mu} ^a$ and the effective scalar fields, $\phi ^a (x)$.  
The missing degrees of freedom are assumed to be the perturbative 
ones which remain after the compression from $A^m _{\mu}$ to
 $\phi ^a (x)$.  These degrees of freedom are handled using standard and 
perturbative techniques. In this paper our focus has been the nonperturbative 
degrees of freedom. 
\par 
We now give a few remarks about the $m^2(\phi^a)^2$ 
and $g^2 a^b_\mu \phi^b a^{c\mu} \phi^c$ terms. The last term, 
$g^2 a^b_\mu \phi^b a^{c\mu} \phi^c$, violates the SU(2) gauge invariance of 
averaged Lagrangian \eqref{sec3-310}. The initial QCD Lagrangian is SU(3) 
gauge invariant, and we attempt to reduce this to an SU(2) invariant one (with the presence
of the Higgs type scalar field this may be a hidden SU(2) symmetry). But the Lagrangian 
\eqref{sec3-310} contains terms which are not SU(2) invariant. What has become
of this desired SU(2) invariance? One possibility is that the averaging, 
$\left\langle Q \left| \ldots \right| Q \right\rangle$, must be taken over all 
different gauge configurations or copies. This is closely connected with the Gribov ambiguity \cite{gribov}
where a gauge is picked, but different gauge configurations satisfy the chosen
condition. In the perturbative regime where one does an expansion in powers of the
coupling constant the Gribov ambiguity is not picked up since the different 
gauge copies are related by a term which is inversely proportional to the coupling constant,
which will therefore not be noticed in a perturbative expansion. In the present case since
we dealing with the nonperturbative regime we must address this averaging over
different gauge copies. We make the assumption that after this averaging over
different gauge copies that all SU(2) gauge invariant terms in \eqref{sec3-310} remain the same, 
but  gauge non-invariant like $g^2 a^b_\mu \phi^b a^{c\mu} \phi^c$ term will go to zero. 
This question is very complicated and will be considered more fully in future 
work. 
\par 
We conclude with a few comments about the generation of a mass term $m^2(\phi^a)^2$. 
In Ref. \cite{vdsin2} the tachyonic mass term from the effective scalar field in the
 SU(2) $\rightarrow$ U(1) reduction was effected via the condensation of ghost fields
that arose from fixing to the Maximal Abelian Gauge \cite{dudal}. 
In the present case this mechanism is not directly applicable, since we
are reducing from a non-Abelian to a smaller non-Abelian group rather than an Abelian
group.  It is possible that a similar ghost condensation mechanism occurs in the present
non-Abelian to non-Abelian reduction. Another option is that the appropriate mass term could 
develop via the Coleman-Weinberg mechanism \cite{coleman1} where radiative corrections 
to the effective scalar field produce a symmetry breaking mass term for $\phi ^a$. Yet another
option for generating the correct mass term for the scalar field is to assume that the {\it SU(2)
gauge field} (not the field $\phi ^a$) develops a positive mass condensate about which the form 
of the ansatz in eq. \eqref{sec3-300} is a fluctuation. For example
\begin{equation}
a^a_0 = 0 \; \;  ; \; \; a^a _i = \frac{m}{g } \delta ^a _i + 
{\tilde a}_i ^a
\label{sec3-320}
\end{equation}
where ${\tilde a}_i ^a$ is a fluctuation about the first term. If
${\tilde a}_i ^a$ takes the form of the monopole ansatz in eq. \eqref{sec3-300}
then one finds that $a^a _{\mu} a^{b \mu} = \frac{m^2}{g^2} \delta ^{ab} + monopole 
\; term$; the cross term goes away due to the symmetry of $\delta _{ia}$ and
antisymmetry of $\epsilon _{aib}$. In this way the last term in \eqref{sec3-310} would 
give rise to the tachyonic mass term 
\begin{equation}
\label{sec3-330}
\frac {g^2}{2} a_{\mu} ^b \phi ^b a^{c \mu} \phi ^c
\rightarrow \frac{m^2}{2} \phi ^a \phi ^a + \frac {g^2}{2} {\tilde a}_{\mu} ^b \phi ^b 
{\tilde a}^{c \mu} \phi ^c .
\end{equation}
This is the most economical method for generating the mass term for $\phi ^a$,
since it turns the unwanted last term of \eqref{sec3-310} into the desired tachyonic
mass. The Lagrangian \eqref{sec3-310} in  terms of ${\tilde a} _{\mu} ^a$ is
almost identical to the Lagrangian in terms of the original $a_{\mu} ^a$, since the two gauge fields
are related by a constant shift. The only additional, different term comes from the 
covariant derivative of the effect scalar field, $\phi ^a$.
A final possibility is that the symmetry breaking mass term can result as 
a consequence of a change of the operator description for strongly nonlinear 
fields as in ref. \cite{dzhun}. 
\par 
In conclusion we would like to emphasize that the problems 
considered here are nonperturbative problems in QCD and therefore have 
the same complexity as confinement. 

\section{Physical Consequences}

The effective Lagrangian arrived at in \eqref{sec3-310} is of the form that 
yields finite energy monopole solutions \cite{thp}. It has been shown 
\cite{coleman} \cite{deser} that classical Yang-Mills
theory do not have finite energy solutions {\it i.e.} there are no
{\it classical} glueballs. The scalar field is crucial 
to having finite energy field configuration. In the present work an effective 
scalar field is introduced via the quantization of the coset field, $A^n _{\mu}$. 
From this one can conclude that even though pure, classical Yang-Mills theory 
does not have finite energy, static field configurations, a quantized Yang-Mills
may support such configurations. Other works have given similar conclusions: 
In \cite{galstov} finite energy solutions were found for the non-Abelian 
Born-Infeld system. In \cite{pavlovsky} it was shown that a modified 
Yang-Mills Lagrangian (with the modifications speculated to come 
from quantization) had finite energy solutions. Thus the monopole solutions 
of the effective Lagrangian \eqref{sec3-310} can be viewed as a type of
color magnetic glueball, since both the SU(2) field, $a_{\mu} ^a$, and the 
effective scalar field, $\phi ^a$, come from a pure SU(3) Yang-Mills theory.

The mass of the monopole solutions is inversely proportional to the square of 
the coupling constant and directly proportional to the mass scale. If the original SU(3)
Lagrangian is associated with the strong interaction then the coupling will be of order
$1$ and the mass scale will be of order $\Lambda \approx 200$ MeV. Thus unlike the 
monopoles in Grand Unified Theories, which have masses greater than $10 ^{15}$ GeV,
the monopoles of the effective Lagrangian \eqref{sec3-310} would have a mass which
would allow them to have physical consequences at low energy scales. 
 
One physical use of these low energy scale monopoles would be to explain 
confinement via the dual superconducting model \cite{mandelstam}. 
In the dual superconductor model of confinement  a condensate of color 
monopoles/antimonopoles is hypothesized to form. This is in analogy to the Cooper 
pair condensate which consists of electrically charged electrons. The color
monopole condensate then exhibits the Meissner effect with respect to color electric 
fields ({\it i.e.} the condensate tries to exclude color electric flux). 
Two color electric charged particles placed in this condensate would then
have their color electric flux squeezed into a thin flux tube or string  between
the color charges \cite{nambu}. This would confine the two color charges, since
as one tried to separate them the energy density would rise linearly with
the distance, rather than falling off with the inverse distance as for a Coulomb
potential. Another possible physical application
for the monopole solutions would be to contribute to the explanation of the
proton spin puzzle. Beginning with the European Muon Collaboration (EMC) \cite{emc}
experiment it was realized that contrary to the simple quark model, 
the spin of the proton comes not only from the spins of the valence quarks, 
but has other contributions. The monopole configurations could provide 
a possible contribution to the proton spin in the form of field angular momentum. 
In refs. \cite{rebbi}  it was demonstrated that the combination of a monopole solution plus 
a particle carrying the ``electric'' charge of the theory gave rise to a field angular momentum
for the composite system.  If the color monopole solutions of  \eqref{sec3-310} arose inside 
the proton they would combine with the color electric, valence  quarks to give gluonic field angular 
momentum contributions to the total proton spin. The field angular of a ``monopole-electric charge"
composite depends on the ``magnetic" and ``electric" charges. For the case in refs.
\cite{rebbi} and also in the present case the field angular momentum would be have a
magnitude of $\hbar$, and would thus be a major contributor to the total spin of the proton.  
\par 
This approach to scalar fields as a condensate of nonperturbative degrees 
of freedom of gauge fields may have interesting applications for gravity where 
scalar fields have various applications: inflation, boson stars, 
non-Abelian black holes and so on. Our approach allows us to speculate that 
these scalar fields are constructed from certain nonperturbative degrees of freedom 
of non-Abelian gauge fields. 
  
\section{Acknowledgments}
VD is very grateful to the ISTC grant KR-677 for the financial support. The work
of DS is partially supported by a 2003 COBASE grant.


\begin{thebibliography}{99}

\bibitem{vdsin1}
V. Dzhunushaliev and D. Singleton, Phys. Rev. \textbf{D65}, 125007 (2002).

\bibitem{vdsin2}
V. Dzhunushaliev and D. Singleton, Mod. Phys. Lett. \textbf{A18}, 955 (2003); hep-th/0210287. 

\bibitem{no} 
H.B. Nielsen and P. Olesen, Nucl. Phys. \textbf{B61}, 45 (1973)

\bibitem{gg} 
H. Georgi and S.L. Glashow, Phys. Rev. Lett. \textbf{22}, 579 (1972)

\bibitem{thp} 
G. 't Hooft, Nucl. Phys. \textbf{B79}, 276 (1974); A.M. Polyakov, JETP
Lett. \textbf{20}, 194 (1974)

\bibitem{emc}
 J. Ashman {\it et. al.}, Phys. Lett. \textbf{B206},
364 (1988); Nucl. Phys. \textbf{B328}, 1 (1989)

\bibitem{coleman} S. Coleman, {\it Classical Lumps and Their Quantum Descendants}
(Erice Lectures, 1975)

\bibitem{deser} S. Deser, Phys. Lett. \textbf{B64}, 463 (1976); H. Pagels,
Phys. Lett. \textbf{B68}, 466 (1977); S. Coleman, Commun. Math. Phys.  \textbf{55}, 113 (1977);  

\bibitem{kondo1} 
Kei-Ichi Kondo, Phys. Rev. \textbf{D57}, 7467 (1998)

\bibitem{heis}
W. Heisenberg, \textit{Introduction to the unified field theory of
elementary particles.}, Max - Planck - Institut f\"ur Physik und
Astrophysik, Interscience Publishers London, New York, Sydney,
1966; W. Heisenberg, Nachr. Akad. Wiss. G\"ottingen, N8,
111(1953); W. Heisenberg, Zs. Naturforsch., \textbf{9a},
292(1954); W. Heisenberg, F. Kortel und H. M\"utter, Zs.
Naturforsch., \textbf{10a}, 425(1955); W. Heisenberg, Zs. f\"ur
Phys., \textbf{144}, 1(1956); P. Askali and W. Heisenberg, Zs.
Naturforsch., \textbf{12a}, 177(1957); W. Heisenberg, Nucl. Phys.,
\textbf{4}, 532(1957); W. Heisenberg, Rev. Mod. Phys., \textbf{29},
269(1957).

\bibitem{giacomo}
A. Di Giacomo, H. G. Dosch, V. I. Shevchenko, Yu. A. Simonov, 
``Field correlators in QCD. Theory and applications'', hep-ph/0007223. 

\bibitem{simonov}
Yu. A. Simonov, ``Selfcoupled equations for the field correlators'', hep-th/9712250. 

\bibitem{gribov} V.N. Gribov, Nucl. Phys. \textbf{  B139}, 1 (1978)

\bibitem{dudal} D. Dudal and H. Verschelde, J. Phys. A {\bf 36}, 8507 (2003); hep-th/0209025

\bibitem{coleman1} S. Coleman and E. Weinberg, Phys. Rev. \textbf{D7}, 1888 (1973)

\bibitem{galstov} D. Gal'stov and R. Kerner, Phys. Rev. Lett. {\bf 84}, 5955 (2000) 

\bibitem{pavlovsky} O. Pavlovsky, Phys. Lett. \textbf{B485} 151 (2000)

\bibitem{rebbi} 
R. Jackiw and C. Rebbi, Phys. Rev. Lett. {\bf 36}, 1116 (1976); P. Hasenfratz and 
G. 't Hooft, Phys. Rev. Lett. {\bf 36}, 1119 (1976)

\bibitem{dzhun} 
V. Dzhunushaliev, Found. Phys. Lett. {\bf 16}, 265 (2003); hep-th/0208088

\bibitem{mandelstam} G. 't Hooft, in {\it High Energy Physics}. ed. A. Zichichi, EPS
International Conference, Palermo (1975); S. Mandelstam, Phys. Rept. {\bf 23}, 245 (1976)

\bibitem{nambu} Y. Nambu, Phys. Rev. {\bf D10}, 4262 (1974)

\end{thebibliography}
\end{document}